\def\pmb#1{\setbox0=\hbox{#1}%
   \kern-.025em\copy0\kern-\wd0
   \kern.05em\copy0\kern-\wd0
   \kern-0.025em\raise.0433em\box0}
\def\gta{\mathrel{{\lower 3pt\hbox{$\mathchar"218$}}\hskip-8pt
   \raise 2pt\hbox{$\mathchar"13E$}}}
\def\lta{\mathrel{{\lower 3pt\hbox{$\mathchar"218$}}\hskip-8pt
   \raise 2pt\hbox{$\mathchar"13C$}}}
\def\half{{\scriptstyle{1\over2}}}
\def\dagg{\phantom{\dagger}}            

\documentclass[twocolumn,showkeys,preprintnumbers,amsmath,amssymb]{revtex4}

\usepackage{graphicx}
\usepackage{dcolumn}
\usepackage{bm}

\begin{document}

\title{A unified theory for the cuprates, iron-based and similar
superconducting systems: \\ 
application for spin and charge excitations in the hole-doped cuprates
}

\author{J. Ashkenazi}
 \email{jashkenazi@miami.edu}
\affiliation{%
Physics Department, University of Miami, P.O. Box 248046, Coral
Gables, FL 33124, U.S.A.\\
}%

\date{\today}

\begin{abstract}
A unified theory for the cuprates and the iron-based superconductors is
derived on the basis of common features in their electronic structures,
including quasi-two-dimensionality, and the large-$U$ nature of the
electron orbitals close to $E_{_{\rm F}}$ (smaller-$U$ hybridized
orbitals reside at bonding and antibonding states {\it away} from
$E_{_{\rm F}}$). Consequently, low-energy excitations are described
in terms of auxiliary particles, representing combinations of
atomic-like electron configurations, rather than electron-like
quasiparticles. The introduction of a Lagrange Bose field is necessary
to enable the treatments of these auxiliary particles as bosons or
fermions. The condensation of the bosons results in static or dynamical
inhomogeneities, and consequently in a commensurate or an incommensurate 
resonance mode. The dynamics of the fermions determines the charge 
transport, and their strong coupling to the Lagrange-field bosons results 
in pairing and superconductivity. The calculated resonance mode in 
hole-doped cuprates agrees with the experimental results, and is shown 
to be correlated with the pairing gap on the Fermi arcs. 
\end{abstract}


\maketitle


The recent discovery of high-temperature superconductivity (SC) in
iron-based compounds, including pnictides \cite{Hosono1, Zhao1, Chu1,
Chu2} and chalcogenides \cite{Wu1} (referred to below as FeSCs),
provides an opportunity to test the validity of high-$T_c$ theories in
correlated-electron systems. Similarly to the cuprates \cite{Honma}, the
FeSCs are derived from an undoped ``parent'' compound which is
generally magnetically ordered \cite{Lynn1, Lynn2, Bao} at low
temperatures and becomes SC under electron- or hole-doping. Also, both
systems are characterized by a layered structure and a
quasi-two-dimensional electronic structure \cite{Singh1, Singh2,
Nekrasov1, Vildosola, Nekrasov2, Singh4, Singh3}. 

A variety of normal-state properties including, {\it e.g.}, the
transport properties ({\i.e.} resistivity, Hall coefficient and
thermoelectric power) of both the cuprates \cite{Takagi,
Kubo,Hwang,Fisher,Tanaka} and the FeSCs \cite{Chu1, Chen1, Chen2} are
characterized by a remarkably similar anomalous behavior. Also, in both
systems the suppression of SC by a high magnetic field results in a
zero-temperature insulator-to-metal transition upon doping \cite{Boeb,
Boebinger}. Even though the pairing symmetry is different in the
cuprates \cite{Tacon, Sawatzky1, Lee} and the FeSCs \cite{Tesanovic,
Takahashi, Moler, Parker}, a resonant spin excitation, characterized by 
wave vectors around those of the magnetic order in the parent compound, 
exists in the SC state of both systems \cite{CRM, ICRM, Christianson}. 

The approximate tetrahedral arrangement of the pnictogen/chalcogen atoms
around the iron atoms in the FeSCs is typical of covalent bonding, and
thus considerable hybridization is expected between orbitals
corresponding to the two atoms. This is confirmed in
electronic-structure calculations \cite{Singh1, Singh2, Nekrasov1,
Vildosola, Nekrasov2, Singh4, Singh3}; however, such hybridization is
found in antibonding and bonding states which lie at least  $\sim
1\;$eV away from the Fermi level ($E_{_{\rm F}}$), while the states at
the close vicinity of $E_{_{\rm F}}$ are non-bonding and of almost a
pure Fe($3d$) nature \cite{Nekrasov1, Vildosola, Nekrasov2, Singh4,
Singh3}. 

Consequently, the intrasite Coulomb and exchange integrals,
corresponding to Wannier functions of the hybridized orbitals of the
entire bands which determine the Fermi-surface (FS), magnetic moments,
{\it etc.}, may be not large in the FeSCs \cite{Anisimov}, resulting in
itineracy and largely reduced magnetic moments \cite{Lynn1, Lynn2, Bao}.
On the other hand, due to the dominantly Fe($3d$) nature of the states
at $E_{_{\rm F}}$, their intrasite integrals are rather large
\cite{Anisimov} and a large-$U$ approach should be applied to study the
physical properties ({\it e.g.} transport and SC) derived from 
low-energy excitations. 

This aspect of the electronic structure of the FeSCs is different from
that of the cuprates, where an entire band around $E_{_{\rm F}}$ is
believed to correspond to the large-$U$ physics \cite{Lee, Anderson},
and an insulating state of large gaps and magnetic moments exists in
the parent compounds. Low-energy carriers are present in the cuprates
due to doping, and since such carriers in {\it both} the cuprates and
the FeSCs correspond to the large-$U$ regime, a unified theory could be
worked out for both of them. This theory should be valid also for other
quasi-two-dimensional SC systems which are close to a magnetic
instability, and have large-$U$ electrons at the vicinity of $E_{_{\rm
F}}$. 

At the basis of this theory stands the observation that SC exists at
stoichiometries where the dynamics of the low-energy carriers dominantly
involves fluctuations between two adjacent occupation numbers ($n$) of
atomic-like configurations ($3d^n$) around the copper or iron atoms. In
the cuprates \cite{Lee, Anderson} these are fluctuations between
effective Cu($3d^9$) (hybridized with O($2p$) orbitals) and
Cu($3d^{10}$) configurations for electron doping, and between effective
Cu($3d^9$) and Cu($3d^8$) (obtained through Zhang-Rice-type
hybridization with O($2p$) orbitals) configurations for hole doping. In
the FeSCs these are fluctuations between Fe($3d^6$) and Fe($3d^7$)
configurations for electron doping, and between Fe($3d^6$) and
Fe($3d^5$) configurations for hole doping. 

Such dynamics of carriers could be treated through the
auxiliary-particle approach \cite{Barnes}. A configuration corresponding
to an occupation number $n$ is denoted by $\alpha(n)$, and a combined
orbital-spin index of an atomic-like electron by $\eta$. For notation
simplicity, let $\alpha(n-1(\eta))$ be the configuration obtained by
removing an $\eta$ electron from $\alpha(n) \ni \eta$. The operator
$a_{i\alpha(n)}^{\dagger}$ creates an auxiliary particle representing
the configuration $\alpha(n)$ at site $i$ (a two-dimensional
approximation is applied of points ${\bf R}_i$ within a planar lattice
which could be defined to contain one Cu or Fe atom per unit cell
\cite{Singh2, Lee1}). 

The creation operators of electrons of spin-orbitals $\eta$ at sites $i$
can be expressed as: 
\begin{equation}
d_{i\eta}^{\dagger} = \sum_n \sum_{\alpha(n)\ni\eta}
a_{i\alpha(n)}^{\dagger} a_{i\alpha(n-1(\eta))}^{\dagg}. 
\label{eq1} 
\end{equation}
They satisfy anticommutation relations of independent fermion operators
under the following conditions: ({\it i}) the consequence of the
large-$U$ approximation that only the contribution of two adjacent
values of $n$ could be considered in rhs of Eq.~(\ref{eq1}) is valid;
({\it ii}) the auxiliary particles created by $a_{i\alpha(n)}^{\dagger}$
are {\it either} bosons for even $n$ and fermions for odd $n$, {\it or}
fermions for even $n$ and bosons for odd $n$; ({\it iii}) the following
constraint is satisfied in every site $i$: 
\begin{equation}
\sum_n \sum_{\alpha(n)} a_{i\alpha(n)}^{\dagger} a_{i\alpha(n)}^{\dagg}
= 1. \label{eq2} 
\end{equation}

As was discussed above, two occupation numbers ($n$) are considered,
including $n_0$ (corresponding to the parent compound), and either 
$n_0 + 1$ (for electron doping) or $n_0 - 1$ (for hole doping). Let us 
denote by $\alpha$, $\beta$ and $\gamma$ the configurations 
corresponding to the occupation numbers $n_0 + 1$, $n_0$ and $n_0 - 1$, 
respectively. Their creation operators at site $i$ are denoted by: 
\begin{equation}
e_{i\alpha}^{\dagger} \equiv a_{i\alpha(n_0 + 1)}^{\dagger}, \ \
s_{i\beta}^{\dagger} \equiv a_{i\alpha(n_0)}^{\dagger}, \ \
h_{i\gamma}^{\dagger} \equiv a_{i\alpha(n_0 - 1)}^{\dagger}. 
\label{eq3} 
\end{equation}
Auxiliary-particles created by $s_{i\beta}^{\dagger}$ are chosen as
bosons, and thus those created by $e_{i\alpha}^{\dagger}$ and
$h_{i\gamma}^{\dagger}$ are fermions. 

The Hamiltonian ${\cal H}$, applied to study low-energy electron
excitations, is based on intrasite one- and two-particle terms, and
intersite one-particle terms. It is expressed in terms of the
auxiliary-particle operators through Eqs.~(\ref{eq1},\ref{eq3}). A
grand-canonical formalism is applied by including in the Hamiltonian
terms corresponding to the chemical potential $\mu$, and to a field of
Lagrange multipliers $\lambda_i$ ($\lambda = \langle \lambda_i \rangle$)
associated with the auxiliary-particles constraint [Eq.~(\ref{eq2})].
The values of $\lambda_i$ and $\mu$ should be determined to yield the
correct charge and constraint in every site. ${\cal H}$ could be,
formally, expressed as (using constraint-preserving term): 
\begin{eqnarray}
{\cal H} &\cong& {\cal H}^s + {\cal H}^e + {\cal H}^h + {\cal H}^{eh} +
\Delta{\cal H}, \label{eq4} \\
{\cal H}^s &=& \sum_{i\beta}(\epsilon^s_{\beta} - \lambda)
s_{i\beta}^{\dagger} s_{i\beta}^{\dagg}, \nonumber \\ 
{\cal H}^e &=& \sum_{i\alpha} \Big\{ (\epsilon^e_{\alpha} - \mu -
\lambda) e_{i\alpha}^{\dagger} e_{i\alpha}^{\dagg} \nonumber \\ &+&
\sum_{j\ne i} \sum_{\alpha^{\prime}\beta\beta^{\prime}} \big[
t_{\beta\alpha^{\prime}}^{\alpha\beta^{\prime}}({\bf R}_i - {\bf R}_j)
s_{i\beta}^{\dagg} s_{j\beta^{\prime}}^{\dagger}
e_{i\alpha^{\prime}}^{\dagger} e_{j\alpha}^{\dagg} + h.c. \big]\Big\},
\nonumber \\ 
{\cal H}^h &=& \sum_{i\gamma} \Big\{ (\epsilon^h_{\gamma} + \mu -
\lambda) h_{i\gamma}^{\dagger} h_{i\gamma}^{\dagg} \nonumber \\ &+&
\sum_{j\ne i} \sum_{\gamma^{\prime}\beta\beta^{\prime}} \big[
t_{\gamma\beta^{\prime}}^{\beta\gamma^{\prime}}({\bf R}_i - {\bf R}_j)
s_{i\beta^{\prime}}^{\dagger} s_{j\beta}^{\dagg} h_{i\gamma}^{\dagg}
h_{j\gamma^{\prime}}^{\dagger} + h.c. \big]\Big\}, \nonumber \\ 
{\cal H}^{eh} &=& \sum_{i\alpha\beta\gamma} \sum_{j\ne i}
\sum_{\beta^{\prime}\ne \beta} \big[
t_{\beta\alpha}^{\beta^{\prime}\gamma}({\bf R}_i - {\bf R}_j)
e_{i\alpha}^{\dagger} h_{j\gamma}^{\dagger} s_{i\beta}^{\dagg}
s_{j\beta^{\prime}}^{\dagg} \nonumber \\ &+& h.c. \big], \nonumber \\ 
\Delta{\cal H} &=& -\sum_{i} (\lambda_i - \lambda) \Big[ \sum_{\alpha}
e_{i\alpha}^{\dagger} e_{i\alpha}^{\dagg} + \sum_{\beta}
s_{i\beta}^{\dagger} s_{i\beta}^{\dagg} \nonumber \\ &+& \sum_{\gamma}
h_{i\gamma}^{\dagger} h_{i\gamma}^{\dagg} \Big]. \nonumber 
\end{eqnarray} 

The $\lambda_i-\lambda$ Lagrange field represents an effective
fluctuating potential which prevents, through $\Delta{\cal H}$,
constraint-violating fluctuations in the auxiliary-particle site
occupation (thus enabling the treatment of atomic-like electron 
configurations as bosons or fermions). The effect of such a fluctuating 
potential on these configurations is analogous to the effect of 
vibrating atoms on electrons. Consequently, similarly to lattice 
dynamics, the quantization of the $\lambda_i-\lambda$ field yields 
bosons. 

In the cuprates one often applies a one-orbital model \cite{Lee,
Anderson}, under which there is one $\alpha$ configuration, corresponding
to a complete Cu($3d^{10}$) shell, one $\gamma$ configuration
corresponding to a Zhang-Rice singlet, and two $\beta$ configurations
corresponding to the spin states of the orbital $\sigma = \uparrow$ and
$\sigma = \downarrow$ (also presented here as $\sigma = \pm 1$). The
present auxiliary-particle method then becomes the ``slave-fermion'' 
method applied in previous works by the author \cite{ashk94, ashk}. 
The parameters appearing in Eq.~(\ref{eq4}) are then simplified to the
intrasite and transfer (hopping) integrals: 
\begin{eqnarray}
&\ & \epsilon^s_{\uparrow} = \epsilon^s_{\downarrow} = \epsilon^d, \ \ \
\epsilon^e_{\alpha} = 2\epsilon^d + U, \ \ \  \epsilon^h_{\gamma} = 0, \
\ \ \nonumber \\ 
&\ & t_{\beta\alpha}^{\alpha\beta}({\bf R}) =
t_{\gamma\beta}^{\beta\gamma}({\bf R}) =
t_{\uparrow\alpha}^{\downarrow\gamma}({\bf R}) =
t_{\downarrow\alpha}^{\uparrow\gamma}({\bf R}) = t({\bf R}). \ \ \ 
\label{eq4a} 
\end{eqnarray} 

In the FeSCs one needs at least three Fe($3d$) orbitals \cite{Singh1,
Singh2, Nekrasov1, Vildosola, Nekrasov2, Singh4, Singh3} (of the $xz$,
$yz$ and $x^2-y^2$ symmetries) to study the electrons at the vicinity of
$E_{_{\rm F}}$, and there are numerous $\alpha$, $\beta$ and $\gamma$
configurations. The parameters appearing in Eq.~(\ref{eq4}) are then
derived from intersite transfer, and intrasite one-particle, Coulomb
and Hund's-rule exchange integrals \cite{Nakamura}. 

Within the large-$U$ approximation, applied in the derivation of ${\cal
H}$, it could be approximated by omitting {\it either} ${\cal H}^e$ and
the $\alpha$ term in $\Delta{\cal H}$, {\it or} ${\cal H}^h$ and the
$\gamma$ term in $\Delta{\cal H}$, and applying a second-order
perturbation expansion in ${\cal H}^{eh}$. This results in corrections
to hopping and intersite exchange terms which are expressed, within a
one-orbital model for the cuprates, as: 
\begin{equation}
\Delta t({\bf R},{\bf R}^{\prime}) \cong - {t({\bf R}) t({\bf
R}^{\prime}) \over U}, \ \ J({\bf R}) \cong {t({\bf R}) t(-{\bf R})
\over U}, \label{eq4b} 
\end{equation}
and Eq.~(\ref{eq4}) is approximately replaced, for hole-doped 
stoichiometries, by:
\begin{eqnarray}
{\cal H} &\cong& \sum_{i} \Big\{( \mu - \lambda) h_{i}^{\dagger}
h_{i}^{\dagg} + \sum_{\sigma} \Big[ (\epsilon^d - \lambda)
s_{i\sigma}^{\dagger} s_{i\sigma}^{\dagg} \nonumber \\ &+& \sum_{j\ne i}
\big[ - \half J({\bf R}_{i} - {\bf R}_{j}) s_{j,-\sigma}^{\dagger}
s_{j,-\sigma}^{\dagg} s_{i\sigma}^{\dagger} s_{i\sigma}^{\dagg}
\nonumber \\ &+& [ t({\bf R}_i - {\bf R}_j) + \sum_{k \ne i,j} \Delta
t({\bf R}_i - {\bf R}_k , {\bf R}_k - {\bf R}_j) \nonumber \\ &\times&
s_{k,-\sigma}^{\dagger} s_{k,-\sigma}^{\dagg} ] s_{i\sigma}^{\dagger}
s_{j\sigma}^{\dagg}  h_i^{\dagg} h_j^{\dagger} \big] \Big] \Big\} +
\Delta{\cal H}. \label{eq4c} 
\end{eqnarray} 

Considering values of $t({\bf R})$ up to third-nearest-neighbor ${\bf
R}$, and of $t({\bf R},{\bf R}^{\prime})$ and $J({\bf R})$ in
Eq.~(\ref{eq4b}) for nearest-neighbor ${\bf R}$ and ${\bf R}^{\prime}$,
yields an expression for ${\cal H}$ in terms of the parameters $t$,
$t^{\prime}$, $t^{\prime\prime}$ and $J$. Values of these parameters for
hole-doped cuprates have been obtained in first-principles calculations
\cite{Andersen1, Andersen, Bansil, Anisimov1}. Explicit expressions for
${\cal H}$ (and its terms discussed further below), derived on the basis
of Eq.~(\ref{eq4c}), will appear elsewhere \cite{Ashkbig}. 

\begin{figure}[t] 
\begin{center}
\includegraphics[width=3.25in]{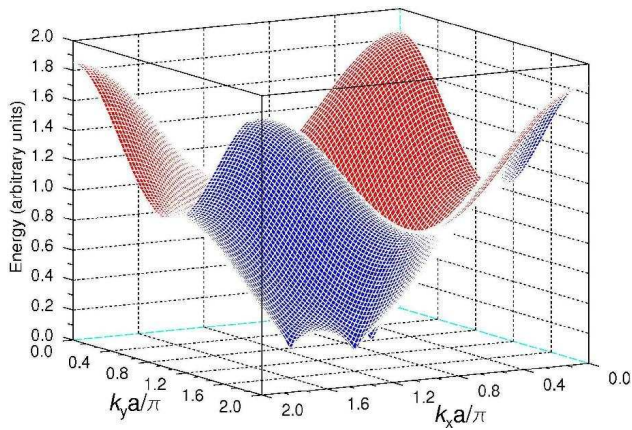}
\end{center}
\caption{A typical lagron spectrum in hole-doped cuprates; the minima 
correspond to striped structures \cite{Tran1, Yamada, Kapitul, Davis}.} 
\label{fig1}
\end{figure}

The Lagrange field bosons are referred to as ``lagrons''. They are soft
at wave vectors corresponding to major fluctuations of spin and orbital
densities. A typical lagron spectrum in hole-doped cuprates is presented
in Fig.~\ref{fig1}; it has soft modes at the points: 
\begin{equation}
{\bf Q}_m = {\bf Q} + \delta {\bf q}_m,\ \text{for} \ m = 1, 2, 3\
\text{or} \ 4, 
\label{eq5} 
\end{equation}
where ${\bf Q} = (\pi/a)({\hat x}+{\hat y})$ is the wave vector of the
antiferromagnetic (AF) order in the parent compounds, and $\delta {\bf
q}_m = \pm \delta q {\hat x} \ \text{or} \ \pm \delta q {\hat y}$ are
modulations around it, corresponding to striped structures
\cite{Tran1, Yamada, Kapitul, Davis}. 

The $s_{i\beta}^{\dagg}$-field bosons are referred to as ``svivons''.
Their Bose condensation is manifested, at low doping levels, in AF
order, in the cuprates \cite{ashk94}, and in a structural distortion and
magnetic order, characterized by a spin-density wave (SDW), in the
FeSCs \cite{Lynn1, Lynn2, Bao}. At higher doping levels the Bose
condensation of svivons is manifested in static or dynamical
inhomogeneities, based on modulations of the low-doping order. 

When svivons are Bose condensed, an $s_{i\beta}^{\dagg}$ field operator
can be expressed as a sum of its ``condensed'' part ({\it i.e.} the
nonzero $\langle s_{i\beta}^{\dagg} \rangle$) and fluctuating part
$s_{i\beta}^{\dagg} - \langle s_{i\beta}^{\dagg} \rangle$. Thus, the
expression of an electron creation operator in term of products of
auxiliary-particle operators, through Eqs.~(\ref{eq1},\ref{eq3}),
includes terms where either $e_{i\alpha}^{\dagger}$ or
$h_{i\gamma}^{\dagg}$ are multiplied by a condensed part of svivon
operators, and terms where they are multiplied by their fluctuating
part. A ``quasi-electron'' (QE) is defined as the fermion created by a
normalized approximation to an electron creation operator, where only
the terms in its expression which include condensed parts of svivon
operators are maintained. 

The QEs represent hypothetical approximate electrons which do not
introduce fluctuations to the inhomogeneities resulting from the Bose
condensation of the svivon field. Since QE states are expanded as
combinations of auxiliary-particle fermion states created by either the
$e_{i\alpha}^{\dagger}$ or the $h_{i\gamma}^{\dagg}$ operators, these
auxiliary-particle states form a basis to the QE states, and could be
referred to as QEs as well. 

Thus, the problem of SC in strongly-interacting electron systems is
treated in terms of an auxiliary space consisting of three types of
coupled ``particles'': ({\it i}) boson svivons which represent
combinations of atomic-like electron configurations of the parent
compounds, and their condensation results in static or dynamical
inhomogeneities; ({\it ii}) fermion QEs which represent combinations of
such configurations with an excess of an electron or a hole over those
of the parent compounds, and their dynamics largely determines charge
transport; ({\it iii}) boson lagrons which represent an effective
fluctuating potential, enabling the treatment of the above
configurations as bosons and fermions. 

Within this auxiliary space the pairing between the fermions through the
exchange of bosons could be rigorously worked out in terms of coupled
{\it independent} fields, in analogy to the electron and phonon fields
within the BCS-Migdal-Eliashberg theory. The strong coupling between
QEs and lagrons, necessary for the constraint [Eq.~(\ref{eq2})] to be
satisfied, results in high pairing temperatures. If the same scenario 
were worked out as the pairing between electrons through the exchange of 
spin or charge fluctuations, generated by {\it the same} system of 
electrons, then two problems would have existed: ({\it i}) it is doubtful 
that such strongly-interacting electrons could be treated as 
quasiparticles; ({\it ii}) the coupled fermion and boson fields are 
{\it not} independent of each other. 

\begin{figure*}[t] 
\begin{center}
\includegraphics[width=3.25in]{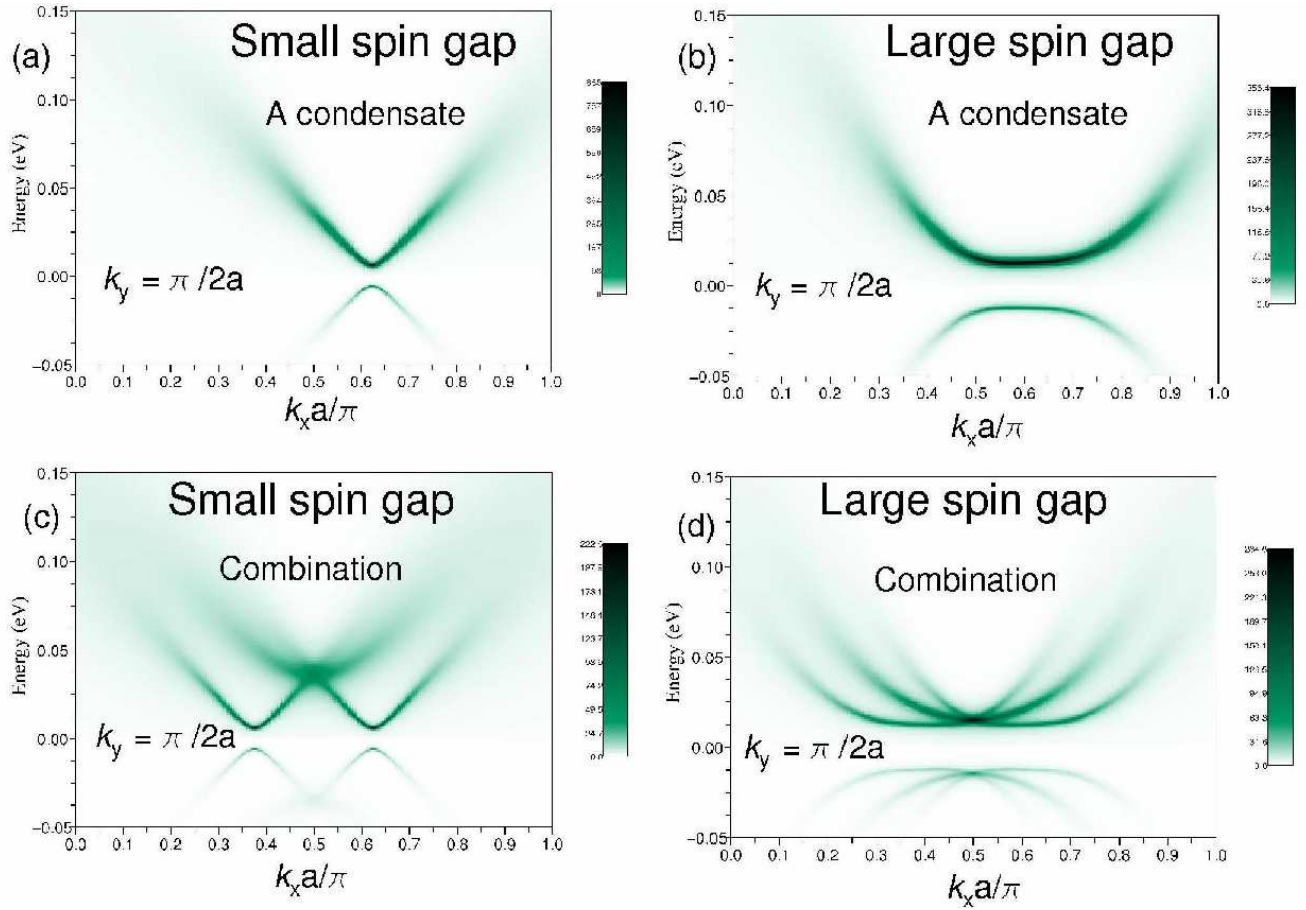}
\end{center}
\caption{The absolute values of the svivon spectral functions below
$T_c$ for two typical hole-doped cuprates of different spin gaps; both
the results for one condensate, and for their average over the four
combined condensates are shown.} 
\label{fig2}
\end{figure*}

Svivon and QE spectra in hole-doped cuprates have been evaluated through
a self-consistent second-order diagrammatic expansion \cite{Ashkbig},
where a mean-field treatment of ${\cal H}$ in Eq.~(\ref{eq4c}) is
applied at the zeroth order. The expansion is carried out on two
Hamiltonian terms. One of them is $\Delta{\cal H}$ which introduces
svivon-lagron and QE-lagron coupling. Vertex corrections to it are
negligible by a phase-space argument, as in Migdal's theorem, since the
dominant contribution of the fluctuating part of the constraint
[Eq.~(\ref{eq2})] comes from a limited ${\bf k}$-space range of the
lagron spectrum around point ${\bf Q}$ (see Fig.~\ref{fig1}). The other
term, ${\cal H}^{\prime}$, introducing QE-svivon coupling, is the
contribution of the fluctuating part of the svivon operators to ${\cal
H}$. It is treated as a perturbation, and approximated through a
first-order expansion of the rhs of Eq.~(\ref{eq4c}) in terms like
$s_{i\sigma}^{\dagg} - \langle s_{i\sigma}^{\dagg} \rangle$ 
\cite{Ashkbig}. 

Lagron spectra of the type presented in Fig.~\ref{fig1} determine
degenerate Bose-condensed svivon states, with energy minima at points
$\pm {\bf Q}_m/2$, for one of the four values of $m$ in Eq.~(\ref{eq5}).
Since there are four inequivalent values of ${\bf Q}/2$ at
$\pm(\pi/2a)({\hat x}\pm{\hat y})$, the number of possible condensates
is eight. In the absence of symmetry-breaking long-range order, the
system is generally in a combination of these states (reflecting
fluctuations between them). Tetragonal symmetry occurs when all the
eight degenerate states are combined, while orthorhombic symmetry
breaking results in the combination of four of the eight states. The
resulting stripe-like inhomogeneities \cite{Tran1, Yamada, Kapitul,
Davis} (which resemble a checkerboard in the combination state) would be
static or dynamical, depending on how close to zero are the spectrum
minima. 

As they occur in Bose fields, the svivon spectral functions are positive
at positive energies, and negative at negative ones. Their absolute
values for two typical cases, of different nonzero spin gaps, in
hole-doped cuprates below $T_c$ (where the low-energy svivon linewidths
are small) are presented in Fig.~\ref{fig2}. Shown are the results for
the svivon condensate with energy minima at $\pm [(\pi/2a) ({\hat x} +
{\hat y}) + \half \delta q {\hat x}]$, and the average of the results
for the four condensates with minima at the vicinity of $\pm (\pi/2a)
({\hat x} + {\hat y})$ (representing their combination), in vertical and
diagonal directions around this point. 

\begin{figure*}[t] 
\begin{center}
\includegraphics[width=3.25in]{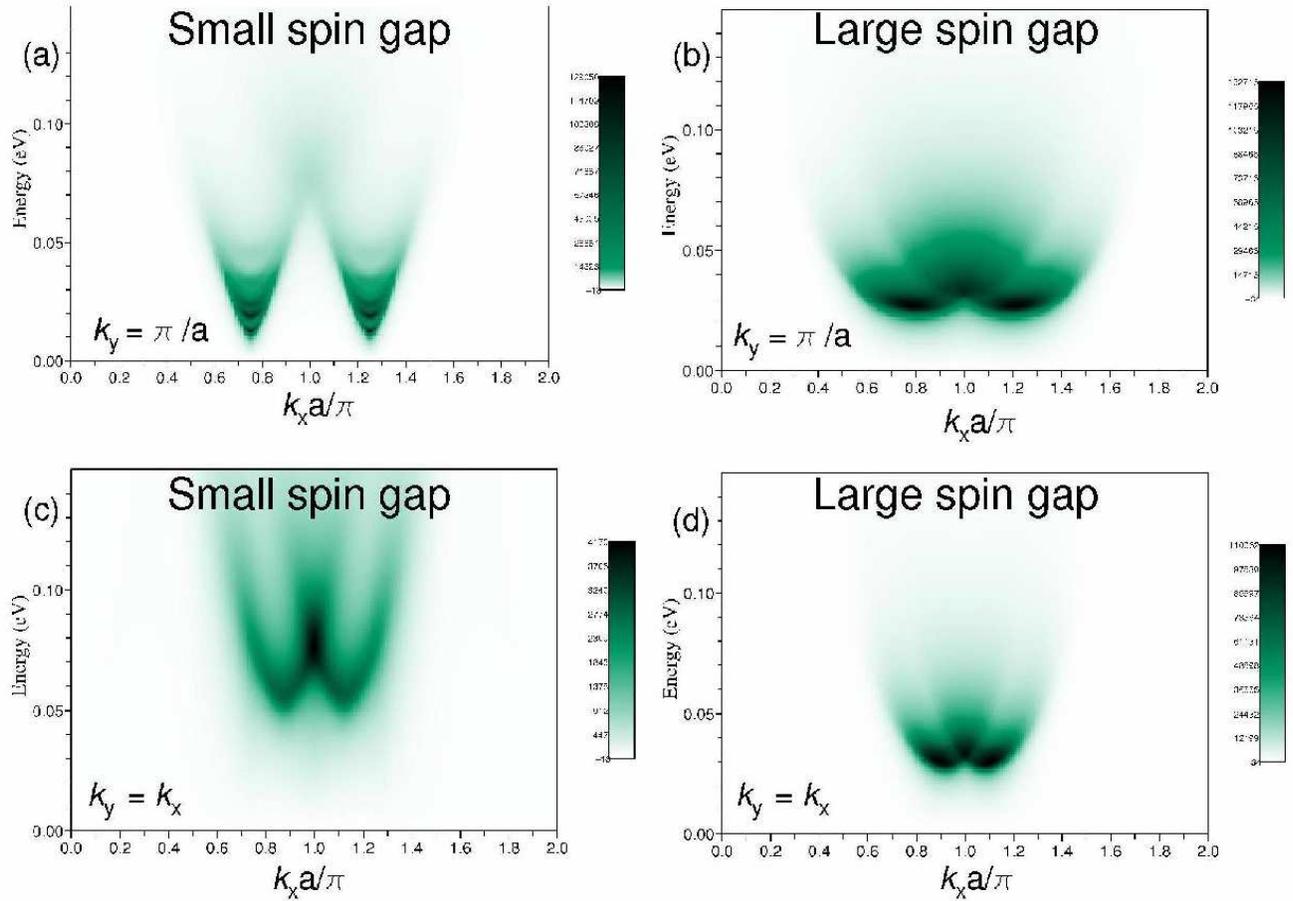}
\end{center}
\caption{The imaginary part of the spin susceptibility, corresponding to
the svivon spectra for hole-doped cuprates below $T_c$ presented in
Fig.~\ref{fig2}; the small- and large-spin-gap results demonstrate,
respectively, the existence of an incommensurate \cite{ICRM} or a
commensurate \cite{CRM} resonance mode.} 
\label{fig3}
\end{figure*}

The QE spectrum of hole-doped cuprates has been evaluated treating the
fluctuations between the combined svivon condensates adiabatically, as
is detailed in a separate paper \cite{AshkQE}. By the definition of
electron creation operators in Eq.~(\ref{eq1}), the electron Green's
functions are obtained at the zeroth-order as sums of products of QE and
svivon Green's functions. This results in the non-Fermi-liquid (non-FL)
scenario of a distribution of convoluted QE-svivon poles. It is shown
\cite{AshkQE} that multiple scattering of QE-svivon pairs introduces to
the electron Green's functions additional FL-like electron poles, and
thus the effect of both types of poles is reflected in various physical
properties. 

The spin susceptibility (SS) of hole-doped cuprates has been evaluated,
under an approximation where only the non-FL convoluted QE-svivon poles
are considered \cite{AshkQE}. Linear-response theory has been applied on
the basis of spin-flip processes, expressed by constraint-preserving
terms of the form $\langle s_{i\sigma}^{\dagger} s_{i,-\sigma}^{\dagg}
s_{j,-\sigma}^{\dagger} s_{j\sigma}^{\dagg} \rangle$, and thus
determined by the svivon spectrum. Results obtained for the imaginary
part of the SS, in vertical and diagonal directions around ${\bf k} =
{\bf Q}$, are presented in Fig.~\ref{fig3}. They correspond to the two
svivon spectra shown in Fig.~\ref{fig2}, and since the svivon-system is
in a combination state, the SS results are averaged over those of the
four combined condensates. These results reproduce those observed in
neutron-scattering measurements in different hole-doped cuprates. The
larger spin-gap results correspond to the ``commensurate resonance mode
(RM)'' \cite{CRM}, and the smaller spin-gap results correspond to the
``incommensurate RM'' \cite{ICRM}. 

\begin{figure*}[t] 
\begin{center}
\includegraphics[width=3.25in]{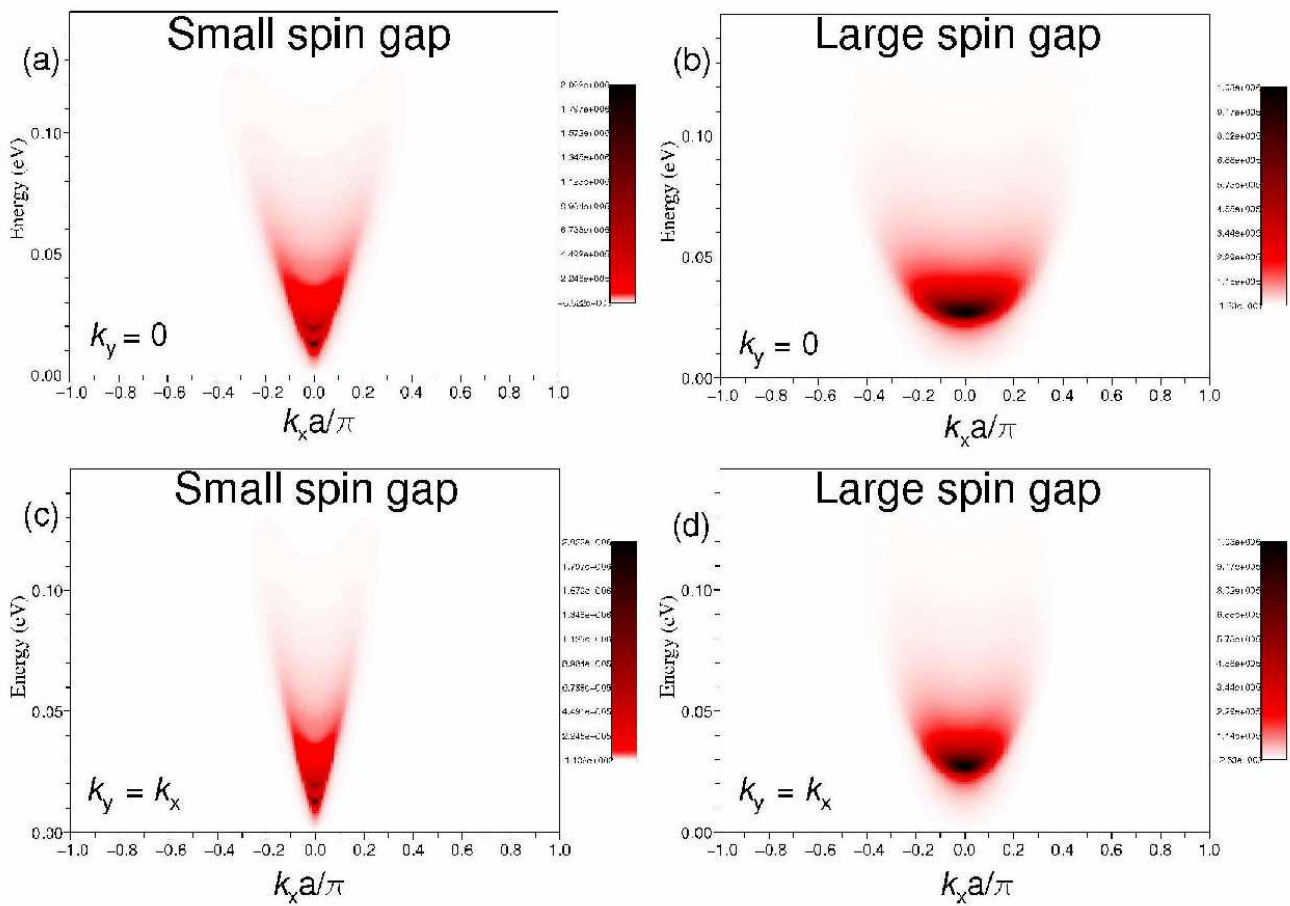}
\end{center}
\caption{The imaginary part of the constraint susceptibility
corresponding to the svivon spectra for hole-doped cuprates below $T_c$
presented in Fig.~\ref{fig2}; the low-energy peaks around ${\bf k}=0$,
approximately, correspond to the integrated energies of the
resonance-mode peaks shown in Fig.~\ref{fig3}; these peaks also,
approximately, correspond through Eq.~(\ref{eq5a}) to the SC gap over
the Fermi arcs (see discussion in the text).} 
\label{fig4}
\end{figure*}

If the constraint is imposed in any two sites $i$ and $j$, the following
equation should be satisfied in these sites in hole-doped cuprates [see
Eqs.~(\ref{eq2},\ref{eq3})]: 
\begin{equation}
\sum_{\sigma\sigma^{\prime}} \langle s_{i\sigma}^{\dagger}
s_{i\sigma}^{\dagg} s_{j\sigma^{\prime}}^{\dagger}
s_{j\sigma^{\prime}}^{\dagg} \rangle \cong \langle h_{i}^{\dagg}
h_{i}^{\dagger} h_{j}^{\dagg} h_{j}^{\dagger} \rangle. 
\label{eq5a} 
\end{equation}
The terms in the lhs of Eq.~(\ref{eq5a}) are formally similar to the
above spin-flip term applied for the derivation of the SS results
presented in Fig.~\ref{fig3}. Thus, a susceptibility-like function,
referred to as the ``constraint susceptibility'' (CS) could be derived
on the basis of either the svivon spectrum, through the lhs of
Eq.~(\ref{eq5a}), or the QE spectrum, through the rhs of
Eq.~(\ref{eq5a}). The results obtained for the CS on the basis of both
spectra should agree with each other in order for the constraint to be 
satisfied, and this condition is the basis for the determination of the 
lagron spectrum, and of their coupling to the svivons and QEs. The CS
represents the response of auxiliary particles, and {\it not} of 
electrons. However, it reflects, under certain conditions, an 
approximation to the response of the system to charge fluctuations 
which could be measured, {\it e.g.}, by Raman spectroscopy \cite{Tacon, 
Gallais1}. 

Results obtained for the imaginary part of the CS, on the basis of the
lhs of Eq.~(\ref{eq5a}), in vertical and diagonal directions around
${\bf k} = 0$, are presented in Fig.~\ref{fig4}. They correspond to the
two svivon spectra shown in Fig.~\ref{fig2}, and evaluated similarly to
the SS results presented in in Fig.~\ref{fig3}. The major feature
observed in CS results is a low-energy peak around ${\bf k}=0$ at
energies which, approximately, correspond to the energies of the ${\bf
k}$-integrated low-energy features of the SS at the vicinity of ${\bf
k}= {\bf Q}$ (thus the incommensurate or commensurate RM). 

Since the {\it same} CS results, as those presented in Fig.~\ref{fig4},
should be obtained also on the basis of the QE spectrum through the rhs
of Eq.~(\ref{eq5a}), and since they correspond to the SC state, the
observed peak at ${\bf k}=0$ should represent some kind of an average
value of the QE gap below $T_c$. As is explained elsewhere
\cite{AshkQE}, this gap has two contributions; one of them originates
from Brillouin zone (BZ) ranges around the antinodal points, where a
narrow peak (of energy $\epsilon=0$ at $T = 0$), lying between two 
humps, splits due to pairing below $T^*$; the other contribution to 
that gap opens below $T_c$ on the Fermi arcs (FAs) around the line 
of nodes. 

In the SC state there are both ``normal'' and ``anomalous''
(pair-correlation) QE Green's functions, and their contributions to the
QE expression for the CS have opposite signs \cite{Ashkbig}. These
contributions cancel each other for ``gap-edge states'', where
$\epsilon=0$, $E= \sqrt{\epsilon^2 + \Delta^2} = \Delta$, and thus the
fraction of both the particle and the hole states within the Bogoliubov
states is $\half [1 \pm \epsilon/E] = \half$. So the QE-spectrum
contributions to the CS peak at ${\bf k}=0$ come from states where
$\epsilon \ne 0$. 

Consequently \cite{AshkQE}, the averaged QE gap which determines the
${\bf k}=0$ CS peak is lowly weighted around the antinodal points, and
represents a value somewhat larger than the averaged QE gap on the FAs.
Since the averaged electron FA gap is also somewhat larger than the QE
FA gap (due to convolution with svivon states), one expects a
correlation between the values of this gap and the ${\bf k}=0$ CS peak,
and as was discussed above (see Figs.~\ref{fig3} and \ref{fig4}), also
with the averaged RM energy. The electron FA gap has been measured
through, {\it e.g.}, the $B_{2g}$ Raman mode, and its value has indeed
been found to be correlated with the RM energy \cite{Tacon, Sawatzky1,
CRM}. A correlation between the energies of the $A_{1g}$ Raman mode and
the RM \cite{Gallais1} has been found to be partial \cite{Tacon1}. The
observed correlation of the FA gap with $\sim 5 k_{_{\rm B}}T_c$
\cite{Tacon, Sawatzky1} is explained elsewhere \cite{AshkQE}. The fact
that the average RM energy is lower when it is incommensurate (see
Fig.~\ref{fig3}) explains the observation that $T_c$ is lower in
cuprates with an incommensurate RM. 

Even though the electronic structure of low-energy states in the FeSCs
is based on more orbitals than in the cuprates, important physical
conclusions could be drawn from one system to the other due to the
formally common Hamiltonian applied for both of them. Within a
two-dimensional approximation, the lagron spectrum of the FeSCs is
expected to differ from that of the cuprates, presented in
Fig.~\ref{fig1}, by replacing the minima positions from satellite points
around ${\bf Q}$, to satellite points around the two possible SDW wave
vectors in the parent compounds: ${\bf Q}_1 = (\pi / a){\hat x}$ and
${\bf Q}_2 = (\pi / a){\hat y}$ \cite{Lynn1, Lynn2} ( or ${\bf Q}_1 =
(\pi / 2a)({\hat x} + {\hat y})$ and ${\bf Q}_2 = (\pi / 2a)({\hat x} -
{\hat y})$ \cite{Bao}), or points close to them. Similarly to the
cuprates \cite {Tran1, Yamada, Kapitul, Davis}, stripe-like
inhomogeneities characterized by modulations due to the differences
between the satellite points and ${\bf Q}_1$ or ${\bf Q}_2$, could exist
also in the FeSCs. 

The svivon spectrum in the FeSCs is expected to have analogous features
to those of the cuprates, presented in Fig.~\ref{fig2}, resulting in a
resonance mode in the vicinity of ${\bf Q}_1$ and ${\bf Q}_2$, below
$T_c$, as has been observed \cite{Christianson}. In a separate paper
\cite{Ashkpair}, it is explained that QE pairing requires a sign
reversal of the order parameter upon a shift of ${\bf Q}$ in the BZ, in
the cuprates, and of ${\bf Q}_1$ or ${\bf Q}_2$ in the FeSCs. Due to
their different FSs, this results in pairing symmetry of an approximate
$d_{x^2-y^2}$ type in the cuprates, and of an $s_{\pm}$ type (thus with
different signs on different FS pockets) in the FeSCs. Thus, it is
predicted that there are no Fermi arcs in the FeSCs, and that their RM
energy is correlated with an averaged value of the SC gap, as has been
observed \cite{Takahashi, Moler, Parker, Christianson}. 

It could be concluded that high-$T_c$ SC occurs in quasi-two-dimensional
strongly-interacting electron systems due to the fact that low-energy
excitations in them are described in terms of auxiliary particles,
representing combinations of atomic-like electron configurations, rather
than electron-like quasiparticles. A Lagrange Bose field which must be
introduced to enable the treatments of these auxiliary particles as
fermions or bosons, serves as the pairing glue between the fermions. 

\begin{acknowledgments}
The author acknowledges the encouragement of Davor Pavuna, and 
constructive discussions with him. Stewart E.~Barnes, Mathieu Le Tacon, 
and many other colleagues and members of the high-$T_c$ community are 
acknowledged for stimulating discussions. Marco Monti is acknowledged 
for his extensive technical assistance. 
\end{acknowledgments}

\end{document}